\documentclass[preprint,12pt]{elsarticle}

\usepackage{amssymb,amsmath}
\usepackage{array, booktabs, multirow}
\usepackage[colorlinks,bookmarksopen,bookmarksnumbered,citecolor=red,urlcolor=red]{hyperref}
\usepackage{pxfonts}
\usepackage[toc,page]{appendix}
\usepackage{lipsum}
\usepackage{hyperref}
\usepackage{cleveref}
\usepackage{dsfont}
\usepackage{multirow}
\usepackage{mwe} 
\usepackage{caption}
\usepackage{epstopdf}
\usepackage[caption=false]{subfig}

\numberwithin{equation}{section}

\journal{impacted journal for publication}

\begin{document}

\begin{frontmatter}



\title{{\bf{Discriminating between and within (semi)continuous classes of both Tweedie and geometric Tweedie models}}}
\author[label1]{Rahma \textsc{Abid}}\ead{Rahma.abid.ch@gmail.com}\address[label1]{Laboratory of Probability and Statistics, University of Sfax, Soukra road B.P.1171, 3038 Sfax, Tunisia}

\author[label2]{C\'{e}lestin C. \textsc{Kokonendji}}\ead{celestin.kokonendji@univ-fcomte.fr}\address[label2]{Laboratoire de math\'ematiques de Besan\c con, Universit\'e Bourgogne Franche-Comt\'e, Besan\c con, France.}

\address{}

\begin{abstract}
In both Tweedie and geometric Tweedie models, the common power parameter $p\notin(0,1)$ works as an automatic distribution selection. It mainly separates two subclasses of semicontinuous ($1<p<2$) and positive continuous ($p\geq 2$) distributions. Our paper centers around exploring  diagnostic tools based on the maximum likelihood ratio test and minimum Kolmogorov-Smirnov distance methods in order to discriminate very close distributions within each subclass of these two models according to values of $p$. Grounded on the unique equality of variation indices, we also discriminate the gamma and geometric gamma distributions with $p=2$ in Tweedie and geometric Tweedie families, respectively. Probabilities of correct selection for several combinations of dispersion parameters, means and sample sizes are examined by simulations. We thus perform a numerical comparison study to assess the discrimination procedures in these subclasses of  two families. Finally, semicontinuous ($1<p\leq 2$) distributions in the broad sense are significantly more distinguishable than the over-varied continuous ($p>2$) ones; and two datasets for illustration purposes are investigated.
\end{abstract}

\end{frontmatter}
\section{Introduction}\label{S1Intro}

Tweedie and geometric Tweedie models provide flexible parametric families of distributions to deal mainly with non-negative right-skewed data and can handle continuous data with probability mass at zero (Tweedie, 1984; J{\o}rgensen and Kokonendji, 2011). They were introduced as tilting exponentials of huge families of stable (e.g., Nolan, 2006) and geometric stable (Klebanov et al., 1984) distributions, respectively. The common power parameter $p\notin (0,1)$, so-called the Tweedie parameter which is one-to-one connected to the common (geometric) stability index $\alpha=(2-p)/(1-p)$, plays an intrinsic role in both models. Indeed, $p$ is an index which distinguishes each distribution within one of each family. See, e.g., Bonat and Kokonendji (2017).

Tweedie distributions are extensively used in statistical modelling and have found a wide range of applications, for instance, in insurance (J{\o}rgensen and Paes De Souza, 1994; Smyth and J{\o}rgensen, 2002), biology (Kendal, 2004, 2007), fisheries research (Foster and Bravington, 2013; Hiroshi, 2008), genetics and medicine (Kendal et al., 2000). Furtheremore, Tweedie family encompasses many special distributions including Gaussian, Poisson, non-central gamma, gamma and inverse Gaussian. The geometric Tweedie family, in turn, arises from geometric sums of Tweedie random variables and may be viewed as the exponential mixture of Tweedie family (e.g., Abid et al., 2019a, 2020). Particular distributions, obviously, stand for the geometric version of the Tweedie ones. In addition, its applications range from ruin probabilities in insurance to waiting times in queueing processes and failures in reliability (Kalashnikov, 1997).

As preliminaries to a discrimination procedure between two distributions, it is necessary that both distributions have common characteristics such as the supports and shapes of densities. More specifically, for both Tweedie and geometric Tweedie families of distributions, we shall also consider zero-mass and variation indices which are recently introduced by Abid et al. (2020) for non-negative random variable $Y$. Recall that the zero-mass index is defined through $\mathrm{ZM}(Y):=\mathbb{P}(Y\leq y)\in [0,1]$ for $y\to 0$. Thus, $\mathrm{ZM}\to\varrho$ when $y\to 0$ indicates a ZM or semicontinuous distribution if $\varrho> 0$ and an absolutely continuous one if $\varrho=0$. As for the variation (or J{\o}rgensen) index expressed by $\mathrm{VI (Y)}=\mathrm{Var Y}/(\mathbb{E}\mathrm{Y})^2  \in (0,\infty)$, it is defined in relation to the standard exponential distribution. Indeed, compared to the well-known  dispersion (or Fisher) index with respect to Poisson for count model (e.g., Kokonendji and Puig, 2018), the VI is viewed as the ratio of the variability of $Y$ to its expected exponential variability which is $(\mathbb{E}Y)^2$. The equi-variation implies no discrepancy between both variabilities. As a matter of fact, $Y$ is said to be over- (equi- and under-varied) compared to exponential with mean $\mathbb{E}Y$ if $\mathrm{VI} > 1$ ($\mathrm{VI} = 1$ and $\mathrm{VI} < 1$), respectively. Scrutinizing both phenomena of ZM and VI, there are very close distributions between and within Tweedie and geometric Tweedie families to be discriminated. Tweedie and geometric Tweedie models are specified and compared in the context of generalized linear models (Kokonendji et al., 2020).

Discriminating between two probability distribution functions was studied by Cox (1961, 1962), Atkinson (1969, 1970), Dyer (1973), and Chen (1980). Dumonceaux and Antle (1973) addressed the problem of discriminating between the log-normal and Weibull distributions. Fearn and Nebenzahl (1991) and Bain and Engelhardt (1980)
tackled the problem of discriminating between the gamma and Weibull distributions. Wiens (1999), Kim et al. (2002), Firth (1988) and
Kundu and Manglick (2005) all handled different aspects of discriminating between the log-normal and gamma distributions. Kundu (2005) discriminated between the normal and Laplace distributions. One can also refer to Algamal (2017), Barreto-Souza and Silva (2015), Kus et al. (2019) and Qaffou and Zoglat (2017) for other distributions. Several authors also considered the discrimination between more than two distributions. Pakyari (2014) discriminated among the generalized exponential, geometric extreme exponential and Weibull distributions. See also Dey and Kundu (2009) for three other distributions. Most of them are based on maximum likelihood ratio test (LRT) and minimum of Kolmogorov-Smirnov distance (KSD). Recently, Rodionov (2018) solved the problem of distinguishing between two close classes of distribution tails. It is expected that a practitioner specifies beforehand the tolerance limits in terms of minimum distances, which are known as tolerance levels among several distribution functions, for discrimination purposes (see, e.g., Gupta and Kundu, 2003). There are certain methodologies to measure the closeness between two distribution functions, such as Kullback-Leibler divergence or Hellinger distance. At this stage, we attemp to challenge a new aspect in statistics in the discrimination between and within close distribution classes of models.

The basic objective of this paper is to discriminate between and within subclasses of both Tweedie and geometric Tweedie models through the use of the maximum LRT and minimum KSD methods.
Sections \ref{S2Tw} and \ref{S3GTw} display some closeness characteristics of the two interested models with the common case of $p=2$. Section \ref{S4GTw} portrays first the proposed discrimination procedures and the estimated probability of correct selection (PCS). Afterwards, it presents some numerical results and reports the challenges involved in the considered discriminations. Section \ref{S6data} is devoted to two analyses of data for illustrative purposes. Eventually, Section \ref{S7GTw} crowns the whole work and provides new perspectives for future research.

\section{Main properties of the Tweedie family}\label{S2Tw}

In this section, some characteristics of continuous and semicontinuous Tweedie models are exhibited. Let $X$ be a random variable distributed as a Tweedie distribution, denoted $Tw_{p}(m,\phi)$. Its density function can be indicated by
\begin{equation}\label{Twee_dens}
f_{Tw_{p}}(x;m,\phi)=a_{p}(x; \phi) \exp [\{ x \psi_{p}(m)  - K_{p}(\psi_{p}(m))\}/\phi]\mathds{1}_{\mathbb{S}_{p}}(x),
\end{equation}
where $\phi>0$ is the dispersion parameter, $p \in(-\infty, 0] \cup[1, \infty)$ is the Tweedie index determining the distribution, $\mathbb{S}_{p}$ is the support of distribution, $a_{p}(x; \phi)$ is the normalizing function to be discussed below, $K_{p}$ is the cumulant function, $\psi_{p}$ is the inverse function of the first derivative $K'_{p}$ and $m=K'_{p}(\theta)$ is the mean of $X$. Note that $K'_{p}(\cdot)$ defines a diffeomorphism between its canonical domain
$\Theta_{p}$ and its image $M_{p}:=K'_{p}(\Theta_{p})$ which is its mean domain. Although the Tweedie densities are not known in a closed form, their cumulant functions are simple. From J{\o}rgensen (1997) we easily deduce that
\begin{equation*}
\psi_{p}(m)=\left\{
\begin{array}{ll}
m^{1-p}/(1-p) & p\neq1\\
\log m & p=1
\end{array}
\right.
\end{equation*}
and, therefore,
\begin{equation*}
K_{p}(\psi_{p}(m))=\left\{
\begin{array}{ll}
m^{2-p}/(2-p) & \mathrm{for}\;\; p\neq2\\
\log m & \mathrm{for}\;\; p=2.
\end{array}
\right.
\end{equation*}

Both sets $\mathbb{S}_{p}$ and $M_{p}$ depend on the value of the power parameter. For $p=0$, $p=1$, $1<p<2$ and $p \geq 2$, the support consists in the real line $\mathbb{R}$, non-negative integers $\mathbb{N}$, non negative real values $[0,\infty)$ and positive values $(0,\infty)$, respectively. The mean domain in these cases is the convex support which is the interior of the closed convex hull of the corresponding $\mathbb{S}_{p}$. Nevertheless, for $p<0$,
one has $\mathbb{S}_{p}=\mathbb{R}$ and $M_{p}=(0,\infty)$. The normalizing function $a_{p}(x; \phi)$ of (\ref{Twee_dens}) cannot be written in a closed form, apart from
the below special cases corresponding to $p=0,1,2,3$. Denoting by $\Gamma(\cdot)$ the classical gamma function and using both $p$ and $\alpha=\alpha(p)$ for simplifying, we successively have
\begin{equation*}
a_{p}(x;\phi)=
\frac{1}{\pi x} \sum_{k=1}^{\infty} \frac{(-x)^k (\alpha \phi)^{k\alpha^{-1}}\Gamma (1+k\alpha^{-1})}{ (\alpha-1)^{\{(\alpha-1)\alpha^{-1} k\}}\Gamma(k+1)} \sin(-k \pi\alpha^{-1})\mathds{1}_{\mathbb{R}}(x)\;\;\mathrm{for}\;\;p<0;
\end{equation*}
\begin{equation*}
a_{p}(x;\phi)=\mathds{1}_{x=0}+\frac{1}{x}\sum_{k=1}^{\infty} \frac{(p-1)^{\alpha k}x^{-k \alpha}}{(2-p)^k \phi^{(1-\alpha)k}\Gamma(-k \alpha)\Gamma(k+1)}\mathds{1}_{x>0}\;\;\mathrm{for}\;\;1<p<2;
\end{equation*}
\begin{equation*}
a_{p}(x;\phi)=
\frac{1}{\pi x}\sum_{k=1}^{\infty} \frac{(p-1)^{\alpha k}\phi^{(\alpha-1)k}\Gamma(1+\alpha k) }{(p-2)^k x^{\alpha k}\Gamma(1+k)} (-1)^k \sin(-k \pi \alpha)\mathds{1}_{x>0} \;\;\mathrm{for}\;\;p>2;
\end{equation*}
and,
\begin{equation*}
a_{p}(x;\phi)=\left\{
\begin{array}{ll}
(2\pi\phi)^{-1/2}\exp(-x^2/2) \mathds{1}_{\mathbb{R}}(x) & \mathrm{for}\;\; p=0\\
\{\Gamma(x+1)\}^{-1} \mathds{1}_{\mathbb{N}}(x) & \mathrm{for}\;\; p=1\\
(1/\phi)^{1/\phi}x^{-1+1/\phi}\{\Gamma(1/\phi)\}^{-1} \mathds{1}_{x>0} & \mathrm{for}\;\; p=2\\
(2\pi x^3\phi)^{-1/2}\exp\{-1/(2x)\} \mathds{1}_{x>0} & \mathrm{for}\;\; p=3.
\end{array}
\right.
\end{equation*}
Departing from (\ref{Twee_dens}), it follows that the zero-mass value of $X\sim Tw_p(m,\phi)$ is deduced by:
\begin{equation*}\label{ZM(Tw)}
\mathrm{ZM}(X)=\exp\left(-\frac{m^{2-p}}{(2-p)\phi} \right)\;\;\mathrm{for}\;\;p\in(1,2);
\end{equation*}
and, therefore, the total mass on the remainder support $(0,\infty)$ is $1-\mathrm{ZM}(X)$ for getting a semicontinuous probability density on $[0,\infty)$ for any $p\in(1,2)$. It is noteworthy that $\mathrm{ZM}=0$ for the gamma distribution with $p=2$ and, the only count distribution of Tweedie family is Poisson with $p=1$.  Dunn and Smyth (2005) elaborated detailed studies on this series and an algorithm to evaluate the Tweedie density function based on expansions series. The algorithm is implemented in the package \texttt{tweedie} (Dunn, 2017) for the statistical software \textsf{R} (\textsf{R} Core Team, 2018) through the function \texttt{dtweedie.series()}. Table \ref{Tab_powEDMs} exhibits some of the subclasses of Tweedie models.
\begin{table}[h]
\begin{center}
{\setlength{\tabcolsep}{0.15cm}
\begin{tabular}{lccll}
\hline\hline
(Geometric) Tweedie models & $\alpha=\alpha(p)$&$p$  & $\mathbb{S}_{p}$&$M_{p}$\\ \hline
(Geometric) Extreme stable &$1<\alpha<2$  &$p<0$ &  $\mathbb{R}$&$(0,\infty)$\\
(Asymmetric Laplace/) Gaussian& $\alpha=2$& $p=0$  & $\mathbb{R}$&$\mathbb{R}$\\
$[$Do not exist$]$ &$\alpha>2$ &$0<p<1$ &\\
(Geometric) Poisson&$\alpha=-\infty$ & $p=1$ &$\mathbb{N}$&$(0,\infty)$ \\
(Geometric) Compound-Poisson-gamma &$\alpha<0$ &$1<p<2$ & $[0,\infty)$&$(0,\infty)$\\
(\emph{Geometric}) \emph{Non-central gamma}&$\alpha=-1$ &$p=3/2$& $[0,\infty)$&$(0,\infty)$\\
\texttt{(Geometric) Gamma}&$\alpha=0$ &$p=2$ & $(0,\infty)$&$(0,\infty)$\\
(Geometric Mittag-Leffler/) Positive stable &$0<\alpha<1$ &$p>2$  & $(0,\infty)$&$(0,\infty)$\\
(\emph{Ressel-Kendall}/) \emph{Inverse Gaussian}&$\alpha=1/2$ &$p=3$ &  $(0,\infty)$&$(0,\infty)$ \\
\hline
\end{tabular}}
\caption{Summary of Tweedie and geometric Tweedie including their common stability index $\alpha=\alpha(p)$, power $p$, support $\mathbb{S}_p$ of distributions and mean domain $M_p$.}
\label{Tab_powEDMs}
    \end{center}
\end{table}

\begin{figure}[h]
\centering
\setlength\fboxsep{6pt}
\setlength\fboxrule{0pt}
    ~~\includegraphics[width=0.8\textwidth]{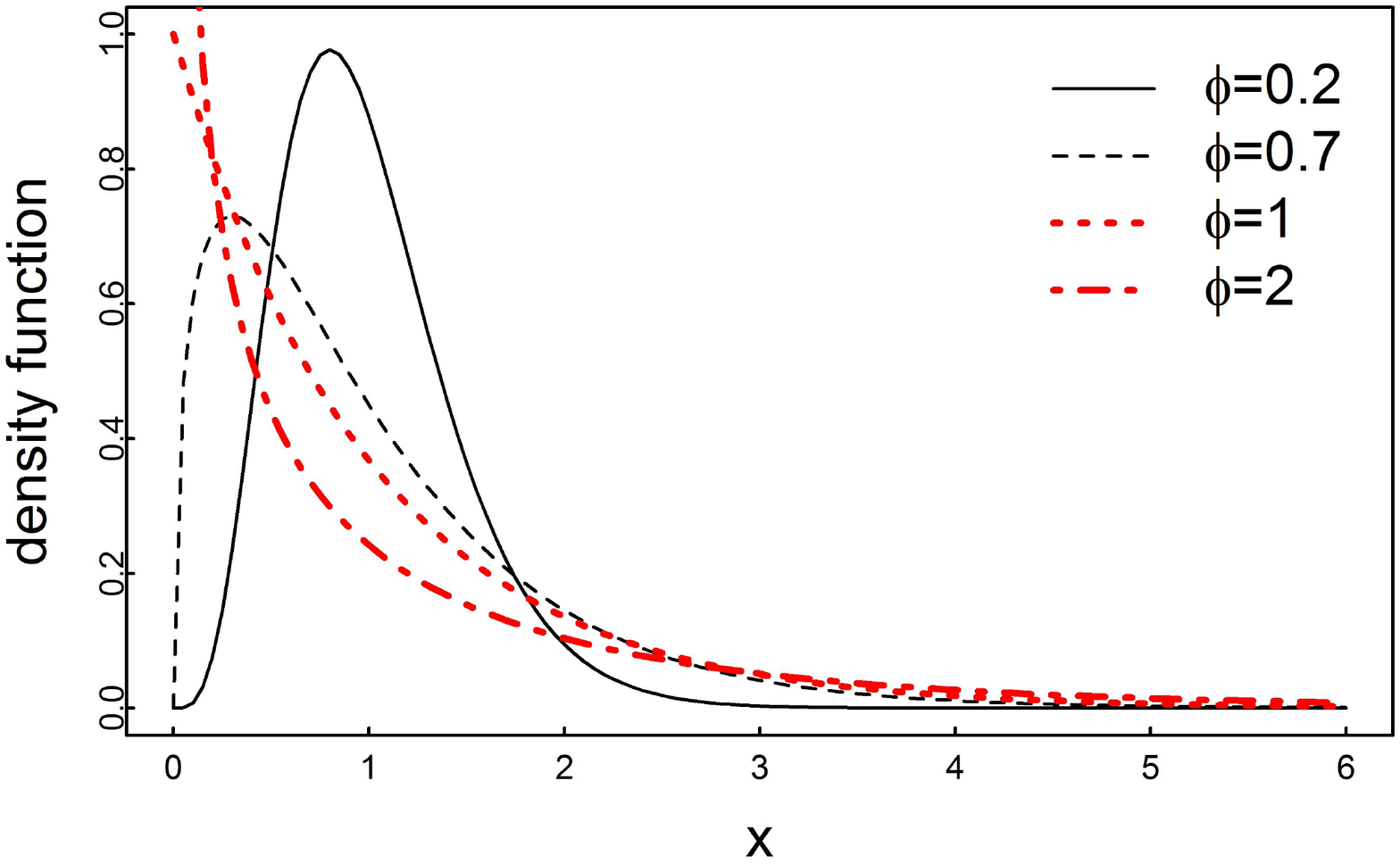}
   \caption{Plots of the density functions of $Tw_{p}(m,\phi)$ with $p=2$, $m=1$ and different values of $\phi$ corresponding to $0<\phi<1$ and $\phi\geq1$.}
\label{Fig_twee}
\end{figure}

\begin{figure}[h]
\includegraphics[width=1.0\textwidth,height=0.3\textheight ]{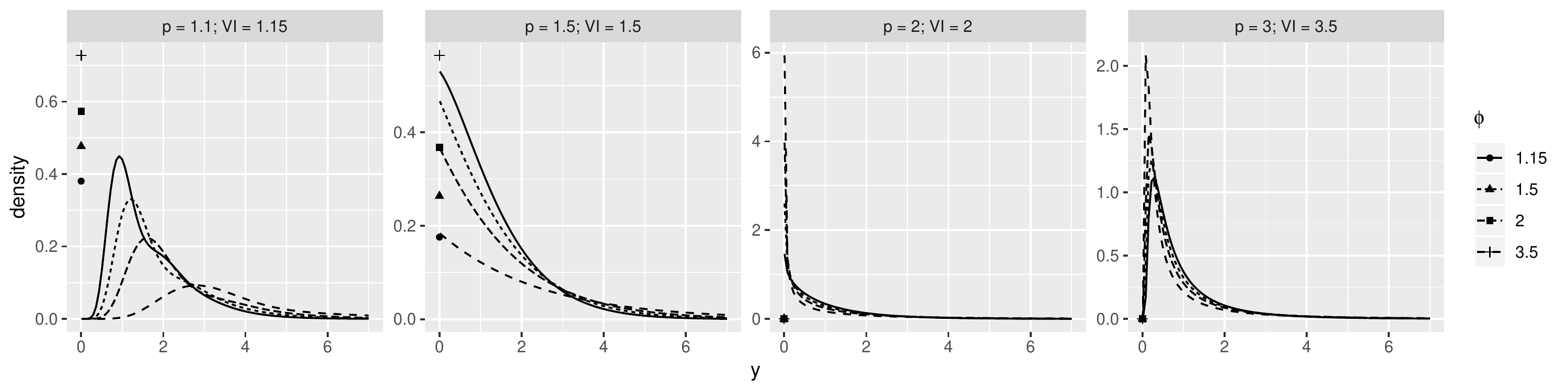}
\caption{Some probability density functions of the Tweedie distribution by parameter configurations. Symbols in the left demonstrate the density at zero.}
\label{Figure_twee4}
\end{figure}

Given the expectation $m$ of $X \sim Tw_{p}(m,\phi)$, its variance is well-known to be $\phi m^{p}$. Thus, one has
\begin{equation}\label{VI_Tw}
\mathrm{VI}(Tw_{p})=\phi m^{p-2}\;\;\left(\:\gtreqqless 1\;\Leftrightarrow\;\phi\gtreqqless m^{2-p}\right).
\end{equation}
Following similar investigations of Abid et al. (2020, Section 4.2 and Figure 1), the dominant behaviors of $\mathrm{VI}(Tw)$ in (\ref{VI_Tw}) appear to be over-variations for all $p\notin(0,1]$ and an equi-variation for $p=2$. This index is new for Tweedie models. The special case of $\mathrm{VI}(Y)=\phi$ in (\ref{VI_Tw}) for the gamma ($p=2$) distribution does not depend on the mean $m$. Figure \ref{Fig_twee} depicts some plots of the Tweedie densities with $p=2$, which illustrates certain similar results that we shall obtain in the next geometric Tweedie models.
Figure \ref{Figure_twee4} displays some plots of Tweedie probability densities such that the dispersion parameter values were chosen in order to have $15\%, 50\%, 100\%$ and $250\%$ much more variability than the reference exponential distribution. These values correspond to $\mathrm{VI} = 1.15, 1.5, 2$ and $2.5$. In addition, these values were combined with different values of the power parameter ($p = 1.1, 1.5, 2$ and $3$) to highlight the flexibility of the distribution to deal with zero-mass ($1<p<2$) and heavy-tailed ($p> 2)$ data.

In order to infer the main properties of Tweedie models before discriminating within them, we shall now investigate their Kullback-Leibler divergences (Kullback and Leibler, 1951). In fact, letting $\varepsilon>0$ the standard difference between two $Tw_{p}(m,\phi)$ according to $p$, the general expression of the Kullback-Leibler divergence between two very close $Tw_{p}(m,\phi)$ and $Tw_{p+\varepsilon}(m,\phi)$ for given $p \notin (0,1]\cup \{2\}$, $\phi>0$ and $m>0$ is indicated as follows:
\begin{eqnarray*}
KL_{p,p+\varepsilon}&=&\int_{-\infty}^{\infty} f_{Tw_{p}}(x;m,\phi) \log \left\{\frac{f_{Tw_{p}}(x;m,\phi)}{f_{Tw_{p+\varepsilon}}(x;m,\phi)}\right\}dx\\
 &=&\int_{-\infty}^{\infty}\left[a_{p}(x;\phi)\exp\left\{ \left(\frac{m^{1-p}}{1-p}x-\frac{m^{2-p}}{2-p}\right)\frac{1}{\phi}\right\}\right]\\
 &&\times\left[\log\left\{ \frac{a_{p}(x,\phi)}{a_{p+\varepsilon}(x,\phi)} \right\}+\frac{x}{\phi}\left(\frac{m^{1-p}}{1-p}-\frac{m^{1-p-\varepsilon}}{1-p-\varepsilon}\right)+\left(\frac{m^{2-p-\varepsilon}}{2-p-\varepsilon}-\frac{m^{2-p}}{2-p}\right)\frac{1}{\phi}\right]dx.
\end{eqnarray*}
Since $KL_{p,p+\varepsilon}$ is analytically tractable, we may resort to the Monte Carlo estimations. Indeed, let ($x_{1},\ldots,x_{n}$) be an $n$-sample of $X \sim Tw_{p}(m,\phi)$. Then, one has
\begin{equation*}
\widehat{KL}_{p,p+\varepsilon}=\frac{1}{n}\sum_{i=1}^{n}\log \left\{ \frac{f_{Tw_{p}}(x_i;m,\phi)}{f_{Tw_{p+\varepsilon}}(x_i;m,\phi)}\right\}.
\end{equation*}

\begin{figure}[h]
 \centering
 \begin{minipage}[c]{\textwidth}
 \centering
        \includegraphics[width=4.5in]{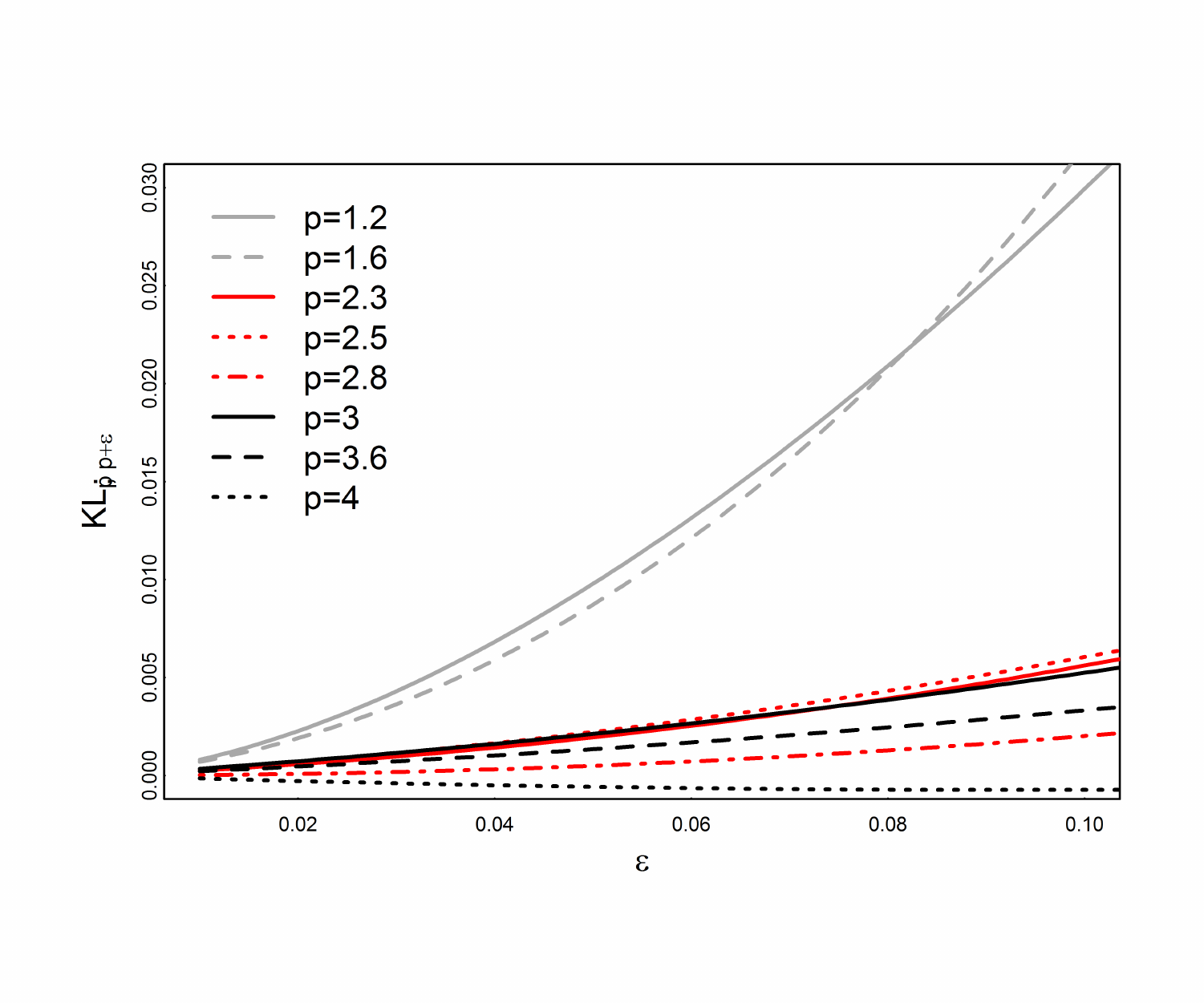}
        \caption{Plots of some estimated Kullback-Leibler divergences between $Tw_{p}$ and $Tw_{p+\varepsilon}$ as functions of $\varepsilon>0$ with different values of $p>1$ and fixed $m=1$ and $\phi=1$.}
       \label{KL}
 \end{minipage}
 \end{figure}

\begin{figure}[h]
\centering
\setlength\fboxsep{3pt}
\setlength\fboxrule{0pt}
    ~~\includegraphics[width=1.0\textwidth]{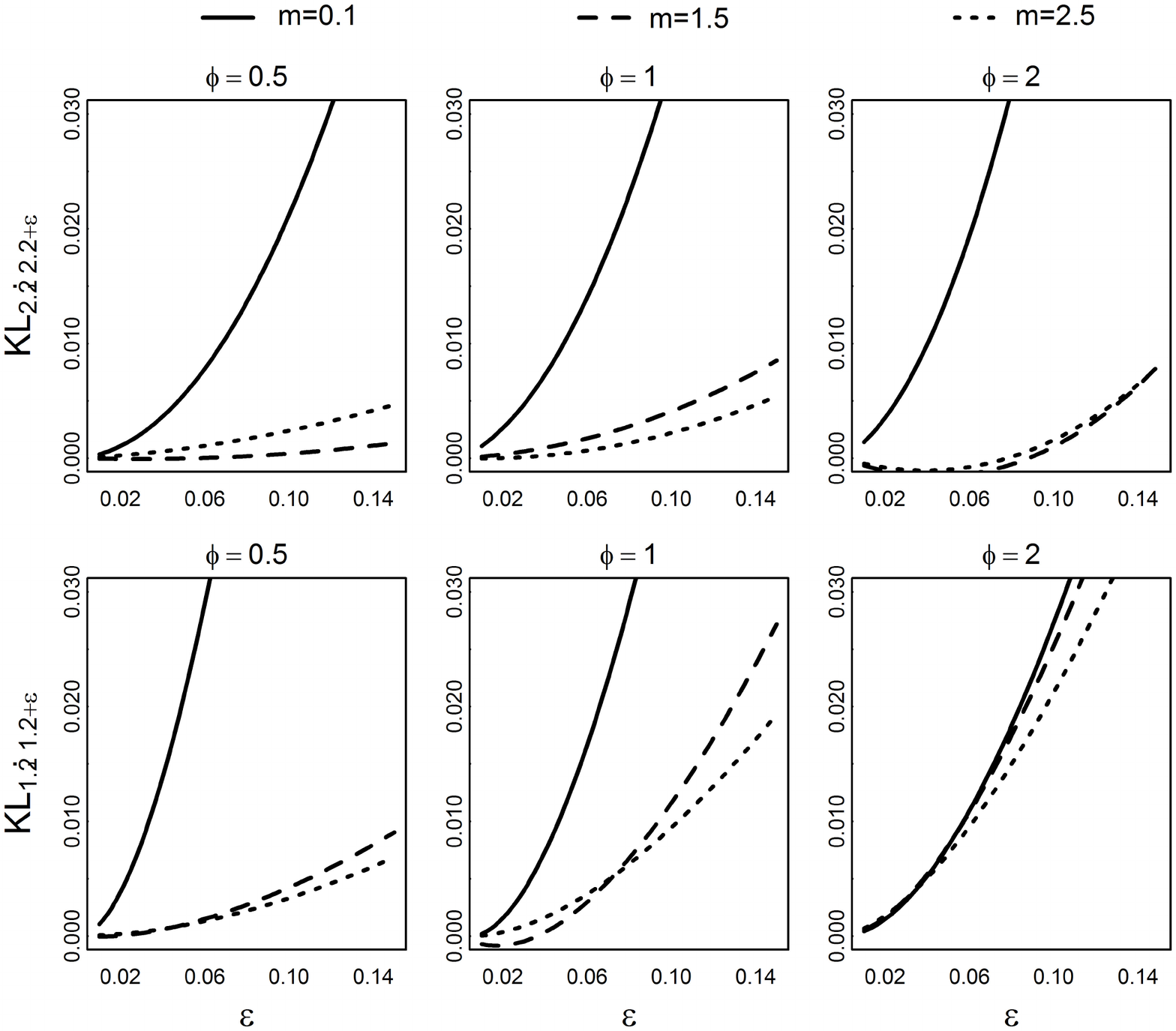}
   \caption{Plots of two representatives of the estimated $KL_{p,p+\varepsilon}$ for $p=1.2$ and $p=2.2$ as functions of $\varepsilon>0$ with different values of $m$ and $\phi$.}
\label{Fig_1_2}
\end{figure}

Figure \ref{KL} displays some estimated $\widehat{KL}_{p,p+\varepsilon}$ as a function of $\varepsilon>0$ for fixed $p>1$, $m=1$ and $\phi=1$. Hence, we clearly observe two groups of similarities through $\widehat{KL}_{p,p+\varepsilon}$ for $p\in(1,2)$ and $p>2$, which are for semicontinuous and continuous subclasses, respectively. It is noteworthy that, for $\varepsilon=0.1$, one gets $\widehat{KL}_{p,p+\varepsilon}=0.030$ for $p\in(1,2)$ and $\widehat{KL}_{p,p+\varepsilon}\leq 0.005$ for $p>2$. Figure \ref{Fig_1_2} depicts one representative from each group of continuous (with $p=2.2$) and semicontinuous (with $p=1.2$) Tweedie models. Basically, if $\varepsilon$ goes to $0$,  then $KL_{p,p+\varepsilon}$ also goes to $0$ for all $p>1$. In such a case, it may be difficult to distinguish these two distributions. However, if $\varepsilon$ increases, then $KL_{p,p+\varepsilon}$ has a higher level for $p\in(1,2)$ and, therefore, the two distributions $Tw_p$ and $Tw_{p+\varepsilon}$ are more dissimilar. Otherwise, when $p\geq 2$, $KL_{p,p+\varepsilon}$ is quasi-invariant to the evolution of $\varepsilon$; and, thus, $Tw_p$ and $Tw_{p+\varepsilon}$ are very close.

\section{Background of the geometric Tweedie family}\label{S3GTw}

Now, we are essentially interested in the continuous and semicontinuous geometric Tweedie models arising from geometric sums of Tweedie variables. Let $Z\sim GTw_{p}(\widetilde{m},\widetilde{\phi})$ be the geometric Tweedie variable with power $p\notin (0,1)$, dispersion $\widetilde{\phi}>0$ and mean $\widetilde{m}$ parameters. Therefore, one has the following representation (e.g., J\o rgensen and Kokonendji, 2011):
\begin{equation*}\label{somme_twee}
Z=\sum_{j=1}^{G}T_{j},
\end{equation*}
where $T_{1},T_{2},\ldots$ are independent and identically distributed (i.i.d.) as a Tweedie distribution $Tw_{p}(m,\phi)$ and $G$ is a geometric random variable, independent of $T_j$, with probability mass function $\mathbb{P}(G=g)=q(1-q)^{g-1}$, for $g=1, 2, \dots$ and $q \in (0,1)$. Moreover, the geometric Tweedie family collapses to exponential mixture representation (see, e.g., Abid et al., 2020, Proposition 2.1) and it is, therefore, expressed by the following hierarchical formulation
\begin{equation*}\label{Yexp-Twp}
\begin{aligned}
X &\sim \mathrm{Exponential}(1) \;\;\;\mathrm{and}\;\;\;
Z|(X=x) & \sim Tw_p(x\widetilde{m},x^{1-p}\widetilde{\phi}).
\end{aligned}
\end{equation*}

The density function of $Z\sim GTw_{p}(\widetilde{m},\widetilde{\phi})$ is deduced from (\ref{Twee_dens}) by
\begin{equation}\label{dens}
f_{GTw_{p}}(z;\widetilde{m},\widetilde{\phi},p) = \int_{0}^{\infty}\exp(-x)f_{Tw_{p}}(z;x\widetilde{m},x^{1-p}\widetilde{\phi})dx,
\end{equation}
which is however not analytically tractable, apart from special cases corresponding to $p\in\{0,1,2,3\}$. Yet, numerical methods allow the density (\ref{dens}) to be accurately and fast evaluated by simulation. The Monte Carlo method provides a very reasonable approximation $\widehat{f}_{GTw_{p}}$ of $f_{GTw_{p}}$, since we have computations of all the Tweedie densities $f_{Tw_{p}}$ through the \textsf{R} function \texttt{dtweedie} (Dunn, 2017). Indeed, let ($x_{1},\ldots,x_{n}$) be an $n$-sample of $X$  following the exponential distribution with unit parameter, then
\begin{equation}\label{dens_approx}
\widehat{f}_{GTw_{p}}(z;\widetilde{m},\widetilde{\phi}) = \frac{1}{n}\sum_{i=1}^{n} f_{Tw_{p}}(z;\widetilde{m}x_{i},x_{i}^{1-p}\widetilde{\phi})\; \xrightarrow{a.s}\;f_{GTw_{p}}(z;\widetilde{m},\widetilde{\phi})\;\mathrm{as}\;n\to\infty,
\end{equation}
where "$\xrightarrow{a.s.}$" stands for the almost surely convergence. Alternatively, we evaluate the integral (\ref{dens}) using the Gauss-Laguerre method and standard \texttt{integrate} function in \textsf{R}. Table \ref{Tab_powEDMs} also presents all types of geometric Tweedie models. From Tweedie models, the only count geometric Tweedie model is the geometric Poisson with $p=1$. The particular case of Ressel-Kendall (see, e.g., Letac and Mora, 1990), which may be called "geometric inverse Gaussian" belongs to the subclass of the geometric Mittag-Leffler ($p > 2$).

Hence, we can deduce an evaluation of the zero-mass $\mathrm{ZM}(Z)$ of $Z\sim GTw_{p}(\widetilde{m},\widetilde{\phi})$ for $p\in(1,2)$ from, either (\ref{dens}) or (\ref{dens_approx}) of the corresponding geometric Tweedie model. For instance, one has $\mathrm{ZM}=0$ for the geometric gamma distribution with $p=2$.  Otherwise, an empirical version of $\mathrm{ZM}(Z)$ can be obtained through
$$
\widetilde{\mathrm{ZM}}(Z)=\frac{1}{n}\sum_{i=1}^n\mathds{1}_{Z_i=z} \;\; \mathrm{for} \;\; z \to 0,
$$
where $(Z_1,\ldots, Z_n)$ is an i.i.d. $n$-sample from $Z\sim GTw_{p}(\widetilde{m},\widetilde{\phi})$.

\begin{figure}[h]
\includegraphics[width=1.0\textwidth,height=0.3\textheight] {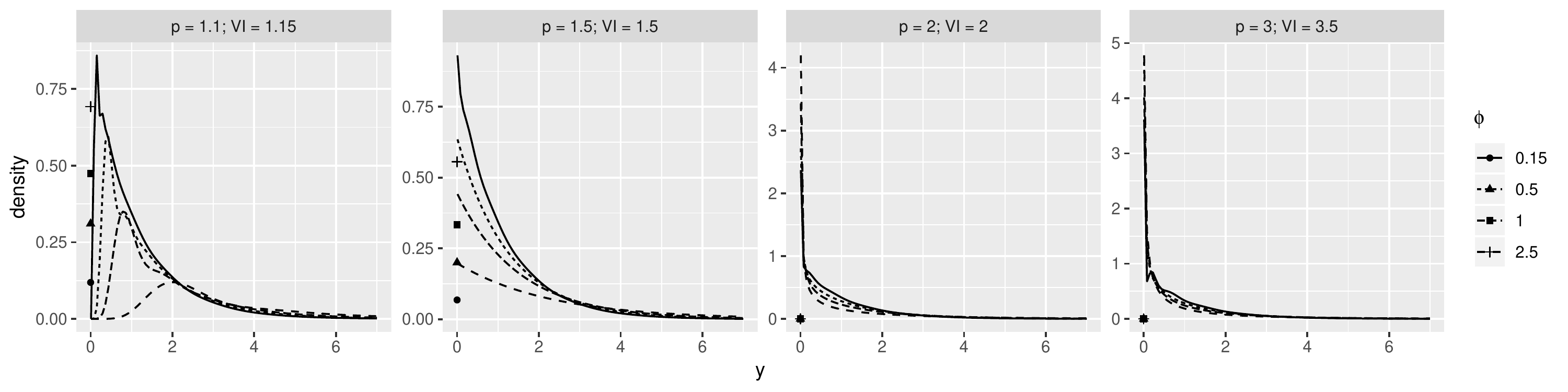}
\caption{Some probability density functions of the geometric Tweedie distribution by parameter
configurations. Symbols in the left indicate the density at zero.}
\label{Figure_twee}
\end{figure}

From the characteristic variance $\widetilde{m}^2 +\widetilde{\phi} \widetilde{m}^{p}$ of $Z\sim GTw_{p}(\widetilde{m},\widetilde{\phi})$, the variation index is expressed by
\begin{equation}\label{VI_gtw}
\mathrm{VI}(GTw_{p})= 1+\widetilde{\phi} \widetilde{m}^{p-2}\;\;\left(\:\gtreqqless 1\;\Leftrightarrow\;\widetilde{\phi}\gtreqqless 0\right).
\end{equation}
Referring to Abid et al. (2020, Section 4.2 and Figure 1) or simply by considering the possibility to get $\widetilde{\phi}$, the dominant features of $\mathrm{VI}(GTw)$ in (\ref{VI_gtw}) of the extended geometric Tweedie models are clearly over-, equi- and under-variations for $p \notin(0,1]$, $p = 2$ and $p\in(-\infty,0]\cup(1,2]$, respectively. However, the associated density function $f_{GTw_{p}}$ does not exist for $\widetilde{\phi} < 0$. Figure \ref{Figure_twee} presents plots of the geometric Tweedie probability density functions. The dispersion parameter values were chosen as in the Tweedie case of Figure \ref{Fig_twee} to highlight the flexibility of the distributions and compare their shapes to the ones from the Tweedie distributions.

It is noteworthy that just like Tweedie models with $p=2$ in (\ref{VI_Tw}), the J\o rgensen (or variation) index $\mathrm{VI}(GTw)$ in (\ref{VI_gtw}) for the particular case $p=2$, corresponding to the geometric gamma distribution,  is equal to $1+\widetilde{\phi}$ and not depending on the mean $\widetilde{m}$. For $p=2$ and given any $\widetilde{m}=m>0$, both variation indexes for Tweedie (\ref{VI_Tw}) and geometric Tweedie (\ref{VI_gtw}) models coincide when their dispersion parameters differ by $+1$ in the sense of geometric Tweedie. More conventionally, one can write $Tw_2(m,\phi)\approx GTw_2(m,1+\phi)$ for $\phi\geq 1$ and any given $m > 0$; which corresponds to two red densities of the Exponential shape in Figure \ref{Fig_twee}.

\begin{figure}[h]
\centering
\setlength\fboxsep{6pt}
\setlength\fboxrule{0pt}
    ~~\includegraphics[width=1\textwidth]{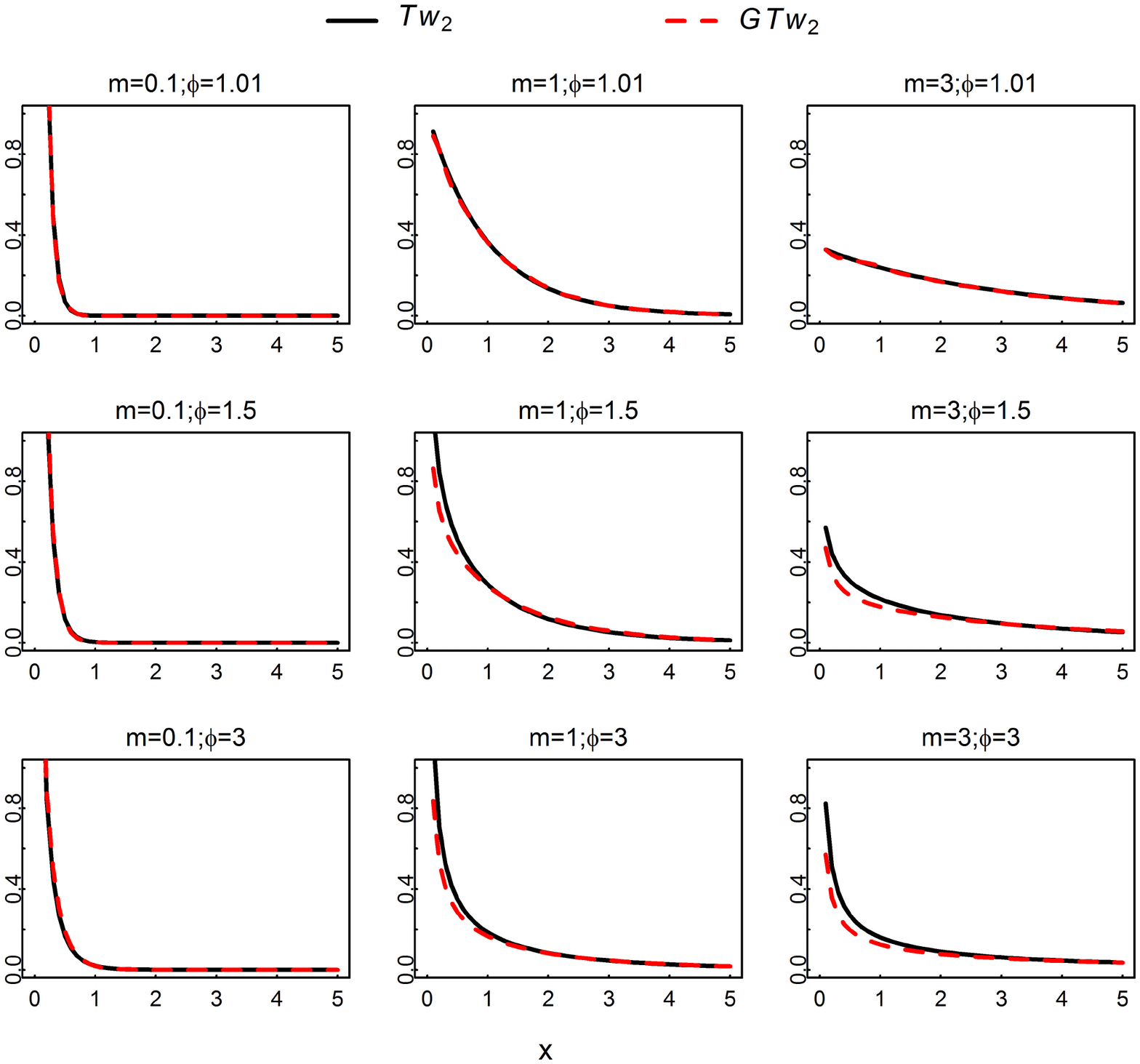}
   \caption{Plots of some density functions with $p=2$ of Tweedie ($\phi>1$) and geometric Tweedie ($\widetilde{\phi}=1-\phi>0$) distributions for given values of $m=\widetilde{m}>0$ and $\phi=1+\widetilde{\phi}>1$.}
\label{Fig}
\end{figure}

A detailed investigation will be carried out below to discriminate these two distributions $Tw_{2}(m,\phi)$ and $GTw_{2}(\widetilde{m},\widetilde{\phi})$. Figure \ref{Fig} displays the densities of Tweedie and geometric Tweedie with  $p=2$ for different values of mean ($m=\widetilde{m}$) and dispersion parameter ($\phi>1$ and $\widetilde{\phi}=1-\phi$). It is clear that both distributions present similar shapes, share the same decreasing behaviour and are quite close, appearing almost indistinguishable. Concerning the corresponding studies of the Kullback-Leibler divergences of the geometric Tweedie family, we here omit them deliberately because they provide similar conclusions for two groups of discriminations within semicontinuous ($1<p<2$) and continuous ($p>2$) subclasses of Tweedie. This is illustrated through Figures \ref{KL} and \ref{Fig_1_2}.

\section{Discrimination procedure and simulation studies}\label{S4GTw}

In this section, two techniques are firstly considered involving the maximum LRT and minimum KSD as optimality criteria to diagnose the appropriate fitting model among two given distributions for a dataset. Next, some simulation studies are performed to compare how the PCSs work for different situations.

Assume that we observe a random sample $Y_{1},Y_{2},\ldots, Y_{n}$ that is supposed to belong to one of the parent distributions $f_{p}(y;m,\phi)$. For fixed $p>1$, the maximum-likelihood of the mean $m$ and dispersion parameter $\phi$ are given, respectively, by
$$\widehat{m}=\frac{1}{n}\sum_{i=1}^{n}Y_{i}\;\;\;\mathrm{and}\;\;\;\widehat{\phi}=\arg\max_{\phi>0}L_{p}(\widehat{m},\phi),$$
where $L_{p}(\widehat{m},\phi)$ is the profile likelihood function calculated at $\widehat{m}$.
The likelihood ratio statistic, also known as the Cox statistic (1961), is defined by
\begin{equation}\label{cox}
LT_{p_{j},p_{j'}}=\log \left(\frac{L_{p_{j}}(\widehat{m}_{j},\widehat{\phi}_{j})}{L_{p_{j'}}(\widehat{m}_{j'},\widehat{\phi}_{j'})}\right),
\end{equation}
The decision rule for discriminating between two distributions having densities $f_{p_{j}}$ and $f_{p_{j'}}$ refers basically to choosing $f_{p_{j}}$ if $LT_{p_{j},p_{j'}} > 0$, and to rejecting  $f_{p_{j}}$ in favor of $f_{p_{j'}}$ otherwise. Notice that, in contrast to the LRT, the KSD test may consider more than two competitive distributions to describe data. The KSD is defined by
\begin{equation}\label{KD}
KS_{p_{j}}=\sup_{-\infty<y<\infty}|\widehat{F}_{p_{j}}(y;\widehat{m}_{j},\widehat{\phi}_{j})-\widetilde{F}(y)|,\;\;\; j \in \{1,\ldots,\ell\},
\end{equation}
with $\ell\geq 2$, $\widehat{F}_{p_{j}}(\cdot;\widehat{m}_{j},\widehat{\phi}_{j})$ the distribution function of $f_{p_{j}}(\cdot;\widehat{m}_{j},\widehat{\phi}_{j})$ and $\widetilde{F}(\cdot)$ the empirical distribution function calculated directly from data. The model index $j_{0}$ with the minimum distance is, therefore, selected as the winning model:
$$j_{0}=\arg\min_{j \in \{1,\dots,\ell\}}KS_{p_{j}}.$$

The performance of the maximum LRT and minimum KSD methods is investigated by the PCSs based on simulations. In practice, we generate $(Y_{n}^{(1)},\ldots,Y_{n}^{(N)})$, where $Y_{n}^{(k)}$ are $k$-random samples of size $n$
that is supposed to belong to $f_{p}$. We repeat both procedures, LRT and KSD, for each $Y_{n}^{(k)}$, $k=1,\dots,N$. The PCS, which corresponds to the proportion of times $f_{p}$, is chosen as the winner model and can be evaluated by:
\begin{equation}\label{PCS}
\widehat{PCS}_{p}=\frac{1}{N}\sum_{k=1}^{N}\mathds{1}\{Y_{n}^{(k)}\; \mathrm{is\; correctly\; classified}\}.
\end{equation}

We will apply both LRT and KSD methods to discriminate between the common and within both Tweedie and geometric Tweedie models. We will compare how the PCS works for different combinations of parameters and sample sizes. All the computations conducted here are performed by  using \textsf{R} (\textsf{R} Core Team, 2018). Of course, the discrimination procedure depends on the quality of the obtained estimates. In addition, it is very crucial to know the closeness between two distribution functions before discriminating them. When a tolerance level is specified (for instance, in terms of the Kullback-Leibler divergence), it means that two distribution functions are not considered to be significantly different if the calculated Kullback-Leibler divergence between them is less than the tolerance level. It is evident that if the corresponding measure is large, then small samples should be enough to discriminate between the two distributions. If the two distribution functions are very close to each other then, naturally, a large sample size is needed to discriminate between them. The models under consideration which are the common ($p=2$), semicontinuous ($1<p<2$) and continuous ($p>2$) of Tweedie and geometric Tweedie merit to be discriminated (see, e.g., Figures \ref{Fig_twee} to \ref{Fig}). They also have the same number of parameters making the discrimination criteria perform quite adequately.

\subsection{Between the gamma and geometric gamma distributions} \label{Commun}

Relying upon their similarities based on the variation indices, and resting on supports and shapes which are previously explored, what is mostly interesting at this level is to discriminate between the gamma $Tw_{2}(m,\phi)$ and geometric gamma $GTw_{2}(\widetilde{m},\widetilde{\phi})$ distributions verifying $\phi=1+\widetilde{\phi}$. Note in passing that, although the Gaussian ($Tw_{0}$) and asymmetric Laplace ($GTw_{0}$) have different VI but the same support and symmetry, Kundu (2005) discriminated between them.

Consider the case when data are coming from gamma $Tw_{2}$ with parameters $m$ and $\phi$. In this case, we considered three values of $\phi = 1.1, 1.5, 3$ combined with three values of $m=0.1, 1.5, 2.5$. For each setting, we generated 100 datasets according to the different sample sizes $n = 20, 40, 60, 80, 100$. The steps for each scenario and for each replica are the following. We needed to find the estimates $\widehat{m}$ and $\widehat{\phi}$ of $Tw_{2}$ and the estimates $\widehat{\widetilde{m}}$ and $\widehat{\widetilde{\phi}}$ of $GTw_{2}$. We used the maximum LRT and the minimum KSD to choose between the two competitive models. In the maximum likelihood ratio procedure, we calculated the likelihood ratio statistic from (\ref{cox}) and the winner model was determined by the decision rule. In the minimum KSD procedure, we calculated the empirical distribution functions of both distributions under consideration using the obtained estimates and we computed the KSD from (\ref{KD}). We estimated the PCS based on simulations as the proportion of times for which the parent distribution $Tw_{2}$ is chosen as the winner model from (\ref{PCS}). The results are reported in Table \ref{TG}.
\begin{table}[h]
\begin{center}
\begin{tabular}{ccccccccc}
    \hline \hline \vspace{-12pt}
      \multirow{ 2}{*}{}&\multirow{ 2}{*}{}&\multicolumn{6}{c}{}\\
       \multicolumn{2}{c}{$Tw_{2}\setminus n$:}&&20&40&60&80&100\\
      $\phi$&$m$&&$GTw_{2}$&$GTw_{2}$&$GTw_{2}$&$GTw_{2}$&$GTw_{2}$\\
      \hline \vspace{-10.5pt}
    \multirow{ 1}{*}{$1.1$}&\multirow{1}{*}{$0.1$}&&&&&&\\
    &&&0.93&0.94&0.98&0.93&0.94\\
    &&&(0.50)&(0.57)&(0.56)&(0.54)&(0.54)\\[-12pt]
    \multirow{ 3}{*}{$1.1$}&\multirow{ 3}{*}{$1.5$}&&&&&&\\
    &&&0.90&0.92&0.95&0.97&1.00\\
    &&&(0.66)&(0.60)&(0.60)&(0.67)&(0.75)\\[-12pt]
    \multirow{ 3}{*}{$1.1$}&\multirow{ 3}{*}{$2.5$}&&&&&&\\
    &&&0.91&0.92&0.95&0.97&0.98\\
    &&&(0.55)&(0.57)&(0.62)&(0.66)&(0.66)\\[-12pt]
    \multirow{ 3}{*}{$1.5$}&\multirow{ 3}{*}{$0.1$}&&&&&&\\
    &&&0.87&0.89&0.90&0.94&0.97\\
    &&&(0.57)&(0.59)&(0.60)&(0.64)&(0.72)\\[-12pt]
    \multirow{ 3}{*}{$1.5$}&\multirow{ 3}{*}{$1.5$}&&&&&&\\
    &&&0.86&0.92&0.93&0.94&0.95\\
    &&&(0.67)&(0.69)&(0.84)&(0.85)&(0.85)\\[-12pt]
    \multirow{ 3}{*}{$1.5$}&\multirow{ 3}{*}{$2.5$}&&&&&&\\
    &&&0.86&0.94&0.93&0.93&0.95\\
    &&&(0.53)&(0.58)&(0.69)&(0.73)&(0.75)\\[-12pt]
   \multirow{ 3}{*}{$3$}&\multirow{ 3}{*}{$0.1$}&&&&&&\\
    &&&0.95&0.98&1.00&1.00&1.00\\
    &&&(0.48)&(0.56)&(0.59)&(0.60)&(0.62)\\[-12pt]
   \multirow{3}{*}{$3$}&\multirow{ 3}{*}{$1.5$}&&&&&&\\
    &&&1.00&1.00&1.00&1.00&1.00\\
    &&&(0.64)&(0.50)&(0.67)&(0.68)&(0.69)\\[-12pt]
    \multirow{3}{*}{$3$}&\multirow{3}{*}{$2.5$}&&&&&&\\
    &&&1.00&1.00&1.00&1.00&1.00\\
    &&&(0.66)&(0.67)&(0.70)&(0.73)&(0.75)\\
 \hline
\end{tabular}
\end{center}
\caption{The estimated PCS when the parent model is gamma $Tw_{2}(m,\phi)$ and the alternative model is geometric gamma $GTw_{2}(\widetilde{m},\widetilde{\phi})$. The numbers in the first row in each box represent the results based on the LRT procedure and the numbers between the parentheses below represent the results based on the KSD procedure.}
\label{TG}
\end{table}

It can be noticed that the PCSs, when calculated by the LRT method, are in general higher than when calculated by the KSD method. It is expected, as far as the distribution functions of Tweedie and geometric Tweedie have tractable analytical form, their empirical distribution functions are used instead. Indeed in these cases, as sample size increases, PCS increases and even when the sample size is 20, results work well. As a matter of fact, the proposed discrimination procedures lead to the selection of the correct model for the given data quite well. Clearly, gamma and geometric gamma are distinguishable.

\subsection{Within (semi)continuous classes of  Tweedie models} \label{sub2}

The parent distribution is $Tw_{p}$ and the alternative distributions are $Tw_{p + \varepsilon}$, with $\varepsilon>0$ such that $Tw_{p}$ and $Tw_{p +\varepsilon}$ have the same type (see Table \ref{Tab_powEDMs}). This part aims to detect the evolution of the discrimination between distributions for each type: $1<p<2$ and $p>2$. We fix two values of $p=1.2, 2.2$ and two degrees of $\varepsilon= 0.1, 0.3$. Such values were chosen in order to cover the maximum range values in a given type of Tweedie. Three values of $\phi = 0.5, 1, 2$ combined with three values of $m=0.1, 1.5, 2.5$ are considered. For each setting, we consider five different sample sizes $n=20, 40, 60, 80,100$ generating $100$ datasets. For each fixed $n$, $\phi$ and $m$, we use the two procedures maximum LRT and minimum KSD in choosing between $Tw_{p}$ and $Tw_{p+\varepsilon}$, first for $\varepsilon=0.1$ and then for $\varepsilon=0.3$. The estimated PCSs are calculated from (\ref{PCS}). Tables \ref{Tw_zero} and \ref{Tw_sup} depict the results when the parent model is $Tw_{1.2}$ and $Tw_{2.2}$, respectively; and, one can observe that the PCS increases as the sample size increases.
\begin{table}[h]
\begin{center}
{\setlength{\tabcolsep}{0.15cm}
\begin{tabular}{cccccccccccccc}
    \hline \hline \vspace{-12pt}
     \multirow{2}{*}{}&\multirow{2}{*}{}&\multicolumn{10}{c}{}\\
      \multicolumn{2}{c}{$Tw_{1.2}\setminus n$:}&\multicolumn{2}{c}{$20$}&\multicolumn{2}{c}{$40$}&\multicolumn{2}{c}{$60$}&\multicolumn{2}{c}{$80$}&\multicolumn{2}{c}{$100$}\\
      $\phi$&$m$&$Tw_{1.3}$ & $Tw_{1.5}$&$Tw_{1.3}$ & $Tw_{1.5}$&$Tw_{1.3}$ & $Tw_{1.5}$&$Tw_{1.3}$ & $Tw_{1.5}$&$Tw_{1.3}$ & $Tw_{1.5}$\\
    \hline \vspace{-10.5pt}
\multirow{ 1}{*}{$0.5$}&\multirow{ 1}{*}{$0.1$}&&&&&&&&\\
&&0.87&099&0.82&1.00&0.82&1.00&0.84&1.00&0.89&1.00\\
&&(0.66)&(0.80)&(0.61)&(0.81)&(0.67)&(0.82)&(0.72)&(0.83)&(0.65)&(0.83)\\[-12pt]

\multirow{ 3}{*}{$0.5$}&\multirow{ 3}{*}{$1.5$}&&&&&&&&&&\\
&&0.55&0.57&0.64&0.69&0.69&0.80&0.72&0.81&0.77&0.90\\
&&(0.47)&(0.56)&(0.49)&(0.52)&(0.52)&(0.64)&(0.52)&(0.59)&(0.53)&(0.61)\\[-12pt]

\multirow{ 3}{*}{$0.5$}&\multirow{ 3}{*}{$2.5$}&&&&&&&&&&\\
&&0.36&0.40&0.47&0.48&0.54&0.53&0.50&0.51&0.59&0.62\\
&&(0.41)&(0.48)&(0.43)&(0.49)&(0.48)&(0.57)&(0.46)&(0.56)&(0.47)&(0.61)\\[-12pt]

\multirow{ 3}{*}{$1$}&\multirow{ 3}{*}{$0.1$}&&&&&&&&&&\\
&&0.73&0.96&0.74&1.00&0.84&1.00&0.66&1.00&0.85&1.00\\
&&(0.62)&(0.80)&(0.69)&(0.99)&(0.71)&(0.94)&(0.56)&(0.94)&(0.70)&(0.99)\\[-12pt]

\multirow{ 3}{*}{$1$}&\multirow{ 3}{*}{$1.5$}&&&&&&&&&&\\
&&0.82&0.92&0.75&0.96&0.76&0.98&0.81&0.98&0.85&0.99\\
&&(0.65)&(0.83)&(0.48)&(0.84)&(0.62)&(0.84)&(0.65)&(0.94)&(0.72)&(0.96)\\[-12pt]

\multirow{ 3}{*}{$1$}&\multirow{ 3}{*}{$2.5$}&&&&&&&&&&\\
&&0.82&0.91&0.77&0.92&0.77&0.97&0.81&0.99&0.89&1.00\\
&&(0.65)&(0.75)&(0.68)&(0.87)&(0.70)&(0.94)&(0.72)&(0.98)&(0.76)&(0.98)\\[-12pt]

\multirow{ 3}{*}{$2$}&\multirow{ 3}{*}{$0.1$}&&&&&&&&&&\\
&&0.98&0.99&0.99&0.99&0.99&1.00&1.00&1.00&1.00&1.00\\
&&(0.98)&(0.98)&(0.98)&(0.99)&(0.98)&(1.00)&(0.99)&(1.00)&(1.00)&(1.00)\\[-12pt]

\multirow{ 3}{*}{$2$}&\multirow{ 3}{*}{$1.5$}&&&&&&&&&&\\
&&0.96&1.00&0.97&1.00&0.99&1.00&1.00&1.00&1.00&1.00\\
&&(0.81)&(0.97)&(0.88)&(0.98)&(0.93)&(1.00)&(0.97)&(1.00)&(0.97)&(1.00)\\[-12pt]

\multirow{ 3}{*}{$2$}&\multirow{ 3}{*}{$2.5$}&&&&&&&&&&\\
&&0.89&0.98&0.96&1.00&0.94&1.00&0.96&1.00&0.98&1.00\\
&&(0.57)&(0.71)&(0.59)&(0.86)&(0.70)&(0.91)&(0.79)&(0.92)&(0.85)&(0.95)\\
\hline
\end{tabular}}
\end{center}
\caption{The estimated PCS when the parent model is $Tw_{1.2}(m,\phi)$ and the alternative model is either $Tw_{1.3}$ or $Tw_{1.5}$. The numbers in the first row in each box represent the results based on the LRT procedure and the numbers between the parentheses below represent the results based on the KSD procedure.}
\label{Tw_zero}
\end{table}

It is worth mentioning from Table \ref{Tw_zero} that as the dispersion parameter $\phi$ moves away from zero, the PCS increases. It is also inferred that, when the alternative is $Tw_{1.5}$, the PCS is higher than the one when the alternative is $Tw_{1.3}$. It is not surprising because as $\varepsilon$ increases, the distance between $Tw_{1.2+\varepsilon}$ increases and therefore it becomes easier to discriminate both of them.

According to Table \ref{Tw_sup} and for small values of $\phi$, we do not have a significant preference over $Tw_{2.2}$ and $Tw_{2.3}$. Indeed, in this case and even with a large sample size, we detect low values of PCS when the alternative model is $Tw_{2.5}$. This refers essentially to the power parameter $p>2$ of Tweedie models which do not allow dissimilarity. This property is significantly reversed for higher values of $\phi$.
\begin{table}[h]
\begin{center}
{\setlength{\tabcolsep}{0.15cm}
\begin{tabular}{cccccccccccccc}
    \hline \hline \vspace{-12pt}
     \multirow{2}{*}{}&\multirow{2}{*}{}&\multicolumn{10}{c}{}\\
      \multicolumn{2}{c}{$Tw_{2.2}\setminus n$:}&\multicolumn{2}{c}{$20$}&\multicolumn{2}{c}{$40$}&\multicolumn{2}{c}{$60$}&\multicolumn{2}{c}{$80$}&\multicolumn{2}{c}{$100$}\\
      $\phi$&$m$&$Tw_{2.3}$ & $Tw_{2.5}$&$Tw_{2.3}$ & $Tw_{2.5}$&$Tw_{2.3}$ & $Tw_{2.5}$&$Tw_{2.3}$ & $Tw_{2.5}$&$Tw_{2.3}$ & $Tw_{2.5}$\\
    \hline \vspace{-10.5pt}
 \multirow{1}{*}{$0.5$}&\multirow{1}{*}{$0.1$}&&&&&&&&\\
&&0.51&0.37&0.52&0.56&0.52&0.57&0.53&0.59&0.58&0.59\\
&&(0.43)&(0.49)&(0.47)&(0.50)&(0.46)&(0.55)&(0.50)&(0.56)&(0.54)&(0.56)\\[-12pt]
\multirow{ 3}{*}{$0.5$}&\multirow{ 3}{*}{$1.5$}&&&&&&&&&&\\
&&0.53&0.49&0.58&0.56&0.52&0.66&0.51&0.68&0.52&0.78\\
&&(0.54)&(0.50)&(0.50)&(0.67)&(0.48)&(0.52)&(0.54)&(0.60)&(0.54)&(0.66)\\[-12pt]
\multirow{ 3}{*}{$0.5$}&\multirow{ 3}{*}{$2.5$}&&&&&&&&&&\\
&&0.50&0.55&0.51&0.61&0.44&0.62&0.56&0.69&0.57&0.77\\
&&(0.46)&(0.55)&(0.48)&(0.63)&(0.53)&(0.62)&(0.56)&(0.68)&(0.58)&(0.69)\\[-12pt]
\multirow{ 3}{*}{$1$}&\multirow{ 3}{*}{$0.1$}&&&&&&&&&&\\
&&0.54&0.46&0.45&0.67&0.42&0.70&0.58&0.75&0.59&0.80\\
&&(0.31)&(0.52)&(0.39)&(0.62)&(0.45)&(0.69)&(0.46)&(0.72)&(0.48)&(0.78)\\[-12pt]
\multirow{ 3}{*}{$1$}&\multirow{ 3}{*}{$1.5$}&&&&&&&&&&\\
&&0.45&0.49&0.53&0.64&0.47&0.69&0.50&0.70&0.52&0.82\\
&&(0.51)&(0.60)&(0.52)&(0.65)&(0.54)&(0.65)&(0.53)&(0.68)&(0.54)&(0.69)\\[-12pt]
\multirow{ 3}{*}{$1$}&\multirow{ 3}{*}{$2.5$}&&&&&&&&&&\\
&&0.44&0.55&0.45&0.58&0.54&0.77&0.44&0.80&0.36&0.78\\
&&(0.65)&(0.60)&(0.67)&(0.64)&(0.62)&(0.67)&(0.66)&(0.77)&(0.59)&(0.75)\\[-12pt]
\multirow{ 3}{*}{$2$}&\multirow{ 3}{*}{$0.1$}&&&&&&&&&&\\
&&0.48&0.75&0.50&0.83&0.66&0.99&0.68&1.00&0.71&1.00\\
&&(0.42)&(0.55)&(0.47)&(0.60)&(0.52)&(0.68)&(0.53)&(0.78)&(0.55)&(0.81)\\[-12pt]
\multirow{ 3}{*}{$2$}&\multirow{ 3}{*}{$1.5$}&&&&&&&&&&\\
&&0.54&0.69&0.54&0.70&0.55&0.80&0.56&0.85&0.56&0.88\\
&&(0.54)&(0.56)&(0.53)&(0.60)&(0.54)&(0.65)&(0.54)
&(070)&(0.57)&(0.75)\\[-12pt]
\multirow{ 3}{*}{$2$}&\multirow{ 3}{*}{$2.5$}&&&&&&&&&&\\
&&0.50&0.65&0.53&0.66&0.56&0.69&0.57&0.76&0.58&0.80\\
&&(0.49)&(0.55)&(0.50)&(0.56)&(0.51)&(0.58)&(0.53)
&(0.60)&(0.57)&(0.65)\\
\hline
\end{tabular}}
\end{center}
\caption{The estimated PCS when the parent model is $Tw_{2.2}(m,\phi)$ and the alternative model is either $Tw_{2.3}$ or $Tw_{2.5}$. The numbers in the first row in each box represent the results based on the LRT procedure and the numbers between the parentheses below represent the results based on the KSD procedure.}
\label{Tw_sup}
\end{table}

\subsection{Within (semi)continuous classes of  geometric Tweedie models}

In this section, the parent distribution is $GTw_{p}(\widetilde{m},\widetilde{\phi})$ and the alternative one is $GTw_{p+\varepsilon}$, with $\varepsilon>0$ such that $GTw_{p}$ and $GTw_{p + \varepsilon}$ have the same type (see Table \ref{Tab_powEDMs}). Here, we will present the behaviour of the maximum LRT and the minimum KSD for only the semicontinuous subclass with $1<p<2$. Simulation results and conclusion for the continuous ($p>2$) subclass of geometric Tweedie  are similar to those of the corresponding Tweedie in Table \ref{Tw_sup}.

For this reason, same parameters as those used in the previous discrimination within the subclass ($1<p<2$) of the Tweedie models are considered. For each combination of sample size, dispersion and mean parameters, parameters are estimated by the pseudo maximum likelihood procedure and the empirical distribution function of geometric Tweedie is calculated.

Table \ref{GTw_sup} presents the obtained estimates of the PCS under the maximum LRT and the minimum KSD methods. We record a good agreement between the PCSs based on maximum LRT and the minimum KSD methods and as $n$ increases, the probabilities go to $1$ as expected. They also lead to the same conclusion as Tweedie models.
\begin{table}[h]
\begin{center}
{\setlength{\tabcolsep}{0.067cm}
\begin{tabular}{cccccccccccccc}
    \hline \hline \vspace{-12pt}
     \multirow{2}{*}{}&\multirow{2}{*}{}&\multicolumn{10}{c}{}\\
      \multicolumn{2}{c}{$GTw_{1.2}\!\setminus\! n$:}&\multicolumn{2}{c}{$20$}&\multicolumn{2}{c}{$40$}&\multicolumn{2}{c}{$60$}&\multicolumn{2}{c}{$80$}&\multicolumn{2}{c}{$100$}\\
      $\widetilde{\phi}$&$\widetilde{m}$&$GTw_{1.3}$ & $GTw_{1.5}$&$GTw_{1.3}$ & $GTw_{1.5}$&$GTw_{1.3}$ & $GTw_{1.5}$&$GTw_{1.3}$ & $GTw_{1.5}$&$GTw_{1.3}$ & $GTw_{1.5}$\\
    \hline \vspace{-10.5pt}
\multirow{ 1}{*}{$0.5$}&\multirow{ 1}{*}{$0.1$}&&&&&&&&\\
&&0.64&0.91&0.80&0.93&0.79&0.98&0.75&0.99&0.79&0.99\\
&&(0.50)&(0.69)&(0.52)&(0.70)&(0.57)&(0.71)&(0.58)&(0.75)&(0.60)&(0.85)\\[-12pt]
\multirow{ 3}{*}{$0.5$}&\multirow{ 3}{*}{$1.5$}&&&&&&&&&&\\
&&0.69&0.83&0.70&0.92&0.77&0.94&0.63&0.96&0.75&0.97\\
&&(0.56)&(0.60)&(0.65)&(0.69)&(0.64)&(0.70)&(0.47)&(0.73)&(0.57)&(0.75)\\[-12pt]
\multirow{ 3}{*}{$0.5$}&\multirow{ 3}{*}{$2.5$}&&&&&&&&&&\\
&&0.69&0.84&0.70&0.87&0.68&0.91&0.69&0.93&0.69&0.98\\
&&(0.53)&(0.54)&(0.57)&(0.63)&(0.54)&(0.65)&(0.55)&(0.72)&(0.59)&(0.71)\\[-12pt]
\multirow{ 3}{*}{$1$}&\multirow{ 3}{*}{$0.1$}&&&&&&&&&&\\
&&0.80&0.98&0.88&1.00&0.81&1.00&0.83&1.00&0.91&1.00\\
&&(0.51)&(0.83)&(0.69)&(0.87)&(0.55)&(0.93)&(0.57)&(0.94)&(0.72)&(0.97)\\[-12pt]
\multirow{ 3}{*}{$1$}&\multirow{ 3}{*}{$1.5$}&&&&&&&&&&\\
&&0.72&0.93&0.76&0.95&0.85&0.98&0.86&1.00&0.88&1.00\\
&&(0.52)&(0.61)&(0.54)&(0.74)&(0.58)&(0.75)&(0.60)&(0.77)&(0.64)&(0.79)\\[-12pt]
\multirow{ 3}{*}{$1$}&\multirow{ 3}{*}{$2.5$}&&&&&&&&&&\\
&&0.82&0.96&0.81&0.97&0.85&0.98&0.88&1.00&0.90&1.00\\
&&(0.50)&(0.69)&(0.59)&(0.70)&(0.53)&(0.71)&(0.60)&(0.79)&(0.75)&(0.80)\\[-12pt]
\multirow{ 3}{*}{$2$}&\multirow{ 3}{*}{$0.1$}&&&&&&&&&&\\
&&0.85&0.99&0.88&1.00&0.90&1.00&0.90&1.00&0.95&1.00\\
&&(0.63)&(0.83)&(0.65)&(0.87)&(0.67)&(0.89)&(0.68)&(0.90)&(0.70)&(0.91)\\[-12pt]
\multirow{ 3}{*}{$2$}&\multirow{3}{*}{$1.5$}&&&&&&&&&&\\
&&0.95&1.00&0.95&1.00&0.96&1.00&096&1.00&0.99&1.00\\
&&(0.67)&(0.83)&(0.67)&(0.84)&(0.69)&(0.85)&(0.70)&(0.88)&(0.77)&(0.90)\\[-12pt]
\multirow{ 3}{*}{$2$}&\multirow{ 3}{*}{$2.5$}&&&&&&&&&&\\
&&0.93&0.97&0.95&1.00&0.98&1.00&0.98&1.00&0.98&1.00\\
&&(0.60)&(0.68)&(0.62)&(0.70)&(0.63)&(0.81)&(0.67)&(0.85)&(0.70)&(0.90)\\

\hline
\end{tabular}}
\end{center}
\caption{The estimated PCS when the parent model is $GTw_{1.2}(\widetilde{m},\widetilde{\phi})$ and the alternative model is either $GTw_{1.3}$ or $GTw_{1.5}$. The numbers in the first row in each box represent the results based on the LRT procedure and the numbers between the parentheses below represent the results based on the KSD procedure.}
\label{GTw_sup}
\end{table}

\section{Real data analysis}\label{S6data}

In this section, two real datasets are analyzed  for illustrative purposes. Concerning the first dataset, gamma $Tw_{2}$ and geometric gamma $GTw_{2}$ distributions are compared. As for the second one, both semicontinuous ($1<p<2$) subclasses of Tweedie and geometric Tweedie are considered through suggesting different values of the power parameter $p$ to fit both models.

\subsection{Failure times of the air conditioning system}

Data consist of the failure times of the air conditioning system of an airplane (Linhart and Zucchini, 1986). They are 23, 261, 87, 7, 120, 14, 62, 47, 225, 71, 246, 21, 42, 20, 5, 12, 120, 11, 3, 14, 71, 11, 14, 11, 16, 90, 1, 16, 52, 95 and already used by Pakyari (2014). The maximum likelihood estimates of the parameters of $Tw_{2}(m,\phi)$ and $GTw_{2}(\widetilde{m},\widetilde{\phi})$ distributions are calculated as $\widehat{m}=59.60$, $\widehat{\phi}=1.2317$, $\widehat{\widetilde{m}}=59.60$ and $\widehat{\widetilde{\phi}}= 0.2380$. It is noteworthy that, $\widehat{\widetilde{\phi}} \simeq 1- \widehat{\phi}$ as expected.

\begin{figure}[h]
\centering
\setlength\fboxsep{6pt}
\setlength\fboxrule{0pt}
    ~~\includegraphics[width=0.8\textwidth]{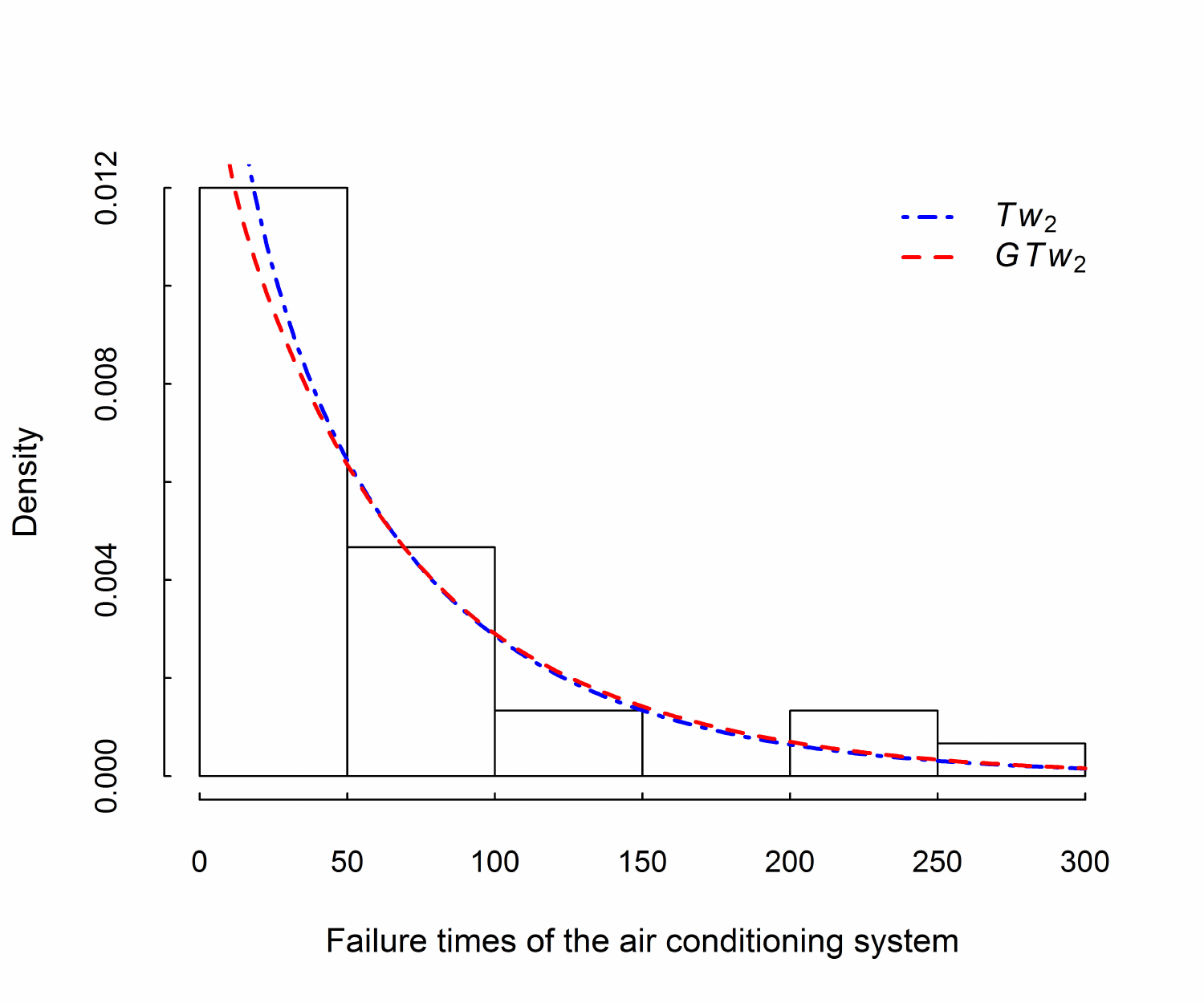}
   \caption{Probability histograms of failure times of the air conditioning system of an airplane and the density functions of the fitted $Tw_{2}$ and $GTw_{2}$ models.}
\label{hist}
\end{figure}

Figure \ref{hist} outlines the histogram and plots of the fitted $Tw_{2}$ and $GTw_{2}$ densities. The log-likelihood functions calculated at the maximum likelihood estimates of the parameters corresponding to $Tw_2$ and $GTw_2$ are obtained  as $ -152.1673$ and $-154.1369$, respectively. Since the value of the log-likelihood functions corresponding to $Tw_{2}$ model is slightly greater than the one of $GTw_{2}$ model, $Tw_{2}$ model would be selected by the LRT method to describe this dataset. The minimum KSD method is also invested.

The KSD between the data and the fitted $Tw_{2}$ distribution function is $0.05213$ whereas the KSD between the data and the fitted $GTw_{2}$ distribution function is $0.09263$. From this perspective, $Tw_{2}$ is again selected with this criterion. Although both criteria suggest $Tw_{2}$, this  choice is not very clear because both fitted $Tw_{2}$ and $GTw_{2}$ distributions are quite close to each other. Probably, the difference in log-likelihood values refers to numerical complexity of $GTw_{2}$ density.

\subsection{Time to failure of pumps}

The second dataset concerns the time to failure of sixty-one cam-driven reciprocating pumps
used in the submarine distillation system over the observation period beginning in May 1987 and ending in December 1993. They are as follows:
1, 2, 2, 2, 3, 4, 5, 8, 10, 11, 13, 13, 15, 17, 18, 26, 27, 29, 10, 14, 14, 18, 21, 24, 28, 31, 34, 38, 41, 61, 15, 21, 23, 26, 33, 41, 43, 43, 56, 10, 25, 39, 42, 48, 52, 24, 26, 34, 43, 44, 49, 51, 37, 40, 0, 0, 0, 0, 0, 0, 0. The source of data is the Navy 3-M system (Dudenhoeffer et al., 1998) and it is also used in Abid et al. (2020) for regression models with covariates.

\begin{figure}[tph]
\centering
\setlength\fboxsep{6pt}
\setlength\fboxrule{0pt}
    ~~\includegraphics[width=0.8\textwidth]{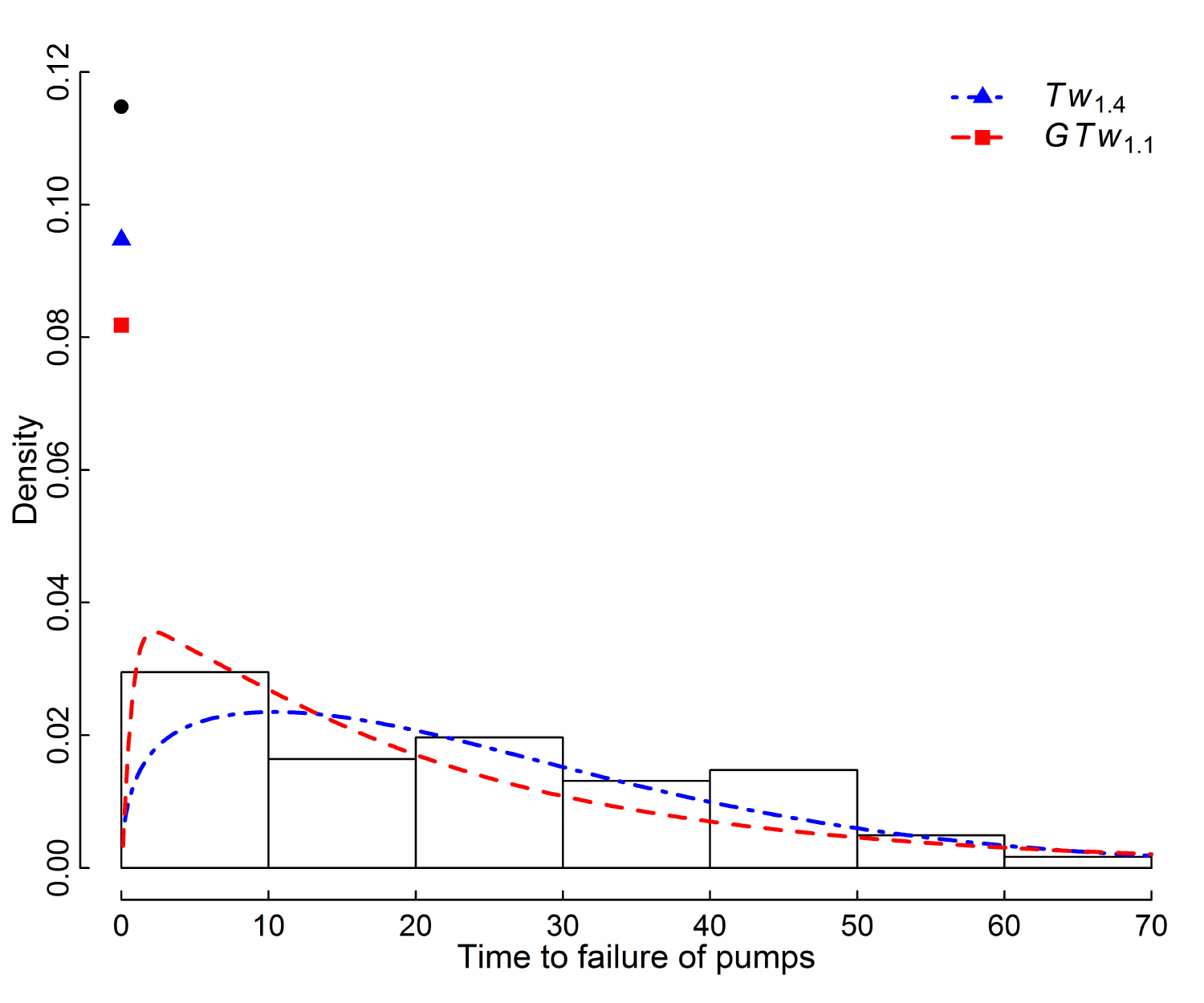}
   \caption{Probability histograms of time to failure of sixty-one cam-driven reciprocating pumps and the density functions of the fitted $Tw_{1.4}$ and $GTw_{1.1}$ models.}
\label{hist_pumps}
\end{figure}

Figure \ref{hist_pumps} displays the histogram of time to failure data with a significant presence of zeros (i.e., $\widehat{\mathrm{ZM}}=0.1148$), which guides us to discriminate among the semicontinuous ($1<p<2$) subclasses of both Tweedie and geometric Tweedie families. For example, nine values of the power parameter between 1.1 and 1.9 are considered. For these data, the mean was found to be $23.0327$. Under the hypothesis that the data come from $Tw_{p}$, the maximum likelihood estimates are computed for each given power parameter.

Table \ref{data_Tw} reveals the estimated dispersion parameters along with the log-likelihood values and KSDs. Based on the log-likelihood values, $Tw_{1.4}$ proves to be the preferred one. Interestingly enough, KSDs suggest to choose $Tw_{1.6}$. However, it is challenging to decide which model between $Tw_{1.3}$, $Tw_{1.4}$, $Tw_{1.5}$, $Tw_{1.6}$ or $Tw_{1.7}$  fits better the corresponding data because the difference in the KSD sense is quite small. Similarly, under the assumption that data come from $GTw_{p}$, the log-likelihood values indicate that $GTw_{1.1}$ is the preferred fitting model. Nevertheless, compared to the log-likelihood value of $GTw_{1.2}$, both distributions fit well these data. As for the KSD values, both $GTw_{1.2}$ and $GTw_{1.3}$ are the best fitting choice for the dataset. Based on the maximum log-likelihood values within these subclasses of Tweedie and geometric Tweedie and for comparison purposes, the two fitted distribution functions $Tw_{1.4}$ and $GTw_{1.1}$ for the dataset are plotted in Figure \ref{hist_pumps}. Another way to select $p$ is to proceed by estimation from data as performed, for instance, by Abid et al. (2020).
\begin{table}[h]
\begin{center}
\begin{tabular}{lccc}
\hline\hline
Models&$\left(\widehat{\widetilde{\phi}}\right)$ $\widehat{\phi}$&Log-lik&KSD\\
 \hline
   $(G)Tw_{1.1}$& (1.8000) 5.7238 & ($-244.7528$) $-266.0258$& (0.0497) 0.1609\\
   $(G)Tw_{1.2}$& (1.5300) 6.1105 & ($-245.7920$) $-250.6113$& (0.0449) 0.1079\\
   $(G)Tw_{1.3}$& (1.2200) 5.4724 & ($-250.2459$) $-246.3995$ &(0.0449) 0.0791\\
   $(G)Tw_{1.4}$& (2.1000) 4.6439 & ($-250.5964$) $-245.9001$& (0.0806) 0.0605\\
   $(G)Tw_{1.5}$& (1.2800) 3.8792 & ($-251.0668$) $-247.2682$& (0.0742) 0.0469\\
   $(G)Tw_{1.6}$& (0.9300) 3.2701 & ($-252.1867$) $-250.1159$ & (0.0672) 0.0379\\
   $(G)Tw_{1.7}$& (0.8600) 2.8560 & ($-253.9838$) $-254.8261$ & (0.0964) 0.0493\\
   $(G)Tw_{1.8}$& (0.6300) 2.7051 & ($-256.0918$) $-262.9210$ & (0.0820) 0.0946\\
   $(G)Tw_{1.9}$& (0.5600) 3.1977 & ($-260.2073$) $-280.2501$ & (0.1076) 0.2092\\
\hline
\end{tabular}
\end{center}
\caption{Estimated dispersion parameters, along with the log-likelihood values (Log-lik) and KSDs for both alternatives $Tw_{p}$ and $GTw_{p}$ models with $1<p<2$. The numbers in the parenthesis represent the results from $GTw_{p}$ models.}
\label{data_Tw}
\end{table}

\section{Concluding remarks}\label{S7GTw}

To this extent, we would assert that we have adopted the maximum LRT and minimum KSD methods to discriminate between gamma and geometric gamma ($p=2$) distributions as well as within semicontinuous ($1<p<2$) and continuous ($p>2$) subclasses of both Tweedie and geometric Tweedie families. The discrimination procedures have been achieved by similar characteristics and appropriate tolerance levels through the Kullback-Liebler divergences. Indeed, LRT method proved to outperform KSD since the PCS clearly allowed the following conclusion: semicontinuous ($1<p\leq 2$) distributions in the broad sense are significantly more distinguishable than the over-varied continuous ($p>2$) ones of both respective families.

We have subsequently illustrated two applications for comparing the best fit in both subclasses of Tweedie and geometric Tweedie models with $p=2$ and $p\in(1,2)$, respectively. As a matter of fact, the last application indirectly demonstrated a possible dissimilarity between semicontinuous distributions of both Tweedie and geometric Tweedie models. However, nothing can be recorded about discrimination between continuous ($p>2$) distributions of the two respective models.

In order to extend the only two count Tweedie and geometric Tweedie distributions, namely Poisson and geometric Poisson with $p=1$, respectively, it is noticed that they are also included in two comparable count classes of the Poisson-Tweedie models (e.g., J{\o}rgensen and Kokonendji, 2016; Kokonendji et al., 2004) and its geometric sums (Abid et al., 2019b), respectively. Both count classes have been also characterized through the use of the common Tweedie parameter $p$ which here belongs to  $\{0\}\cup[1,\infty)$. At this stage of analysis, promising as it may be, our research can be extended and built upon as tackling this area is crucial to fulfill a constructive and fruitful contribution to the count field in terms of further discrimination of additional similar characteristics.

\end{document}